\documentclass[%
reprint,
preprintnumbers,
superscriptaddress,
 amsmath,amssymb,
 aps,
]{revtex4-1}

\usepackage[normalem]{ulem}
\usepackage{graphicx}
\usepackage{dcolumn}
\usepackage{bm}
\usepackage{color}
\usepackage[utf8]{inputenc}
\usepackage{hyperref}
\definecolor{blue}{rgb}{0,0,0.5}
\definecolor{lightblue}{rgb}{0,0,1}
\definecolor{red}{rgb}{0.5,0,0}
\definecolor{lightred}{rgb}{1,0.5,0}
\definecolor{green}{rgb}{0,0.5,0}
\definecolor{darkgreen}{rgb}{0.0,0.3,0.0}
\definecolor{orange}{rgb}{1,0.4,0}
\definecolor{grey}{rgb}{0.5,0.5,0.5}
\definecolor{nicered}{rgb}{0.0,.7,.3}
\definecolor{nicegreen}{rgb}{.1,.5,.1}
\definecolor{darkblue}{rgb}{0,.1,.9}
\hypersetup{
   unicode=false,          
   pdftoolbar=true,        
   pdfmenubar=true,        
   pdffitwindow=false,     
   pdfstartview={FitH},    
   pdftitle={My title},    
   pdfauthor={Author},     
   pdfsubject={Subject},   
   pdfcreator={Creator},   
   pdfproducer={Producer}, 
   pdfkeywords={keyword1} {key2} {key3}, 
   pdfnewwindow=true,      
   colorlinks=true,        
   linkcolor=darkblue,         
   citecolor=nicered,      
   filecolor=magenta,      
   urlcolor=nicered         
}

\begin{document}

\preprint{FERMILAB-PUB-18-336-T}
\preprint{OSU-HEP-18-04}

\title{
Dark Neutrino Portal to Explain MiniBooNE excess
}

\author{Enrico Bertuzzo}
\email[E-mail:]{bertuzzo@if.usp.br}
 \affiliation{Departamento de F\'isica Matem\'atica, Instituto de F\'isica\\
Universidade de S\~ao Paulo, C.P. 66.318, S\~ao Paulo, 05315-970, Brazil}
\author{Sudip Jana}%
 \email[E-mail:]{sudip.jana@okstate.edu}
\affiliation{Department of Physics and Oklahoma Center for High Energy Physics,\\
 Oklahoma State University, Stillwater, OK 74078-3072, USA}%
\affiliation{Theory Department, Fermi National Accelerator Laboratory, P.O. Box 500, 
Batavia, IL 60510, USA}%
\author{Pedro A. N. Machado}
\email[E-mail:]{pmachado@fnal.gov}
\affiliation{Theory Department, Fermi National Accelerator Laboratory, P.O. Box 500, 
Batavia, IL 60510, USA}%
\author{Renata Zukanovich Funchal}
\email[E-mail:]{zukanov@if.usp.br}
 \affiliation{Departamento de F\'isica Matem\'atica, Instituto de F\'isica\\
Universidade de S\~ao Paulo, C.P. 66.318, S\~ao Paulo, 05315-970, Brazil}

\date{\today}

\begin{abstract}
We present a novel framework that provides an explanation to the long-standing excess of
electron-like events in the MiniBooNE experiment at Fermilab. We
suggest a new dark sector containing a dark neutrino and a dark
gauge boson, both with masses between a few tens and a few
hundreds of MeV.  Dark neutrinos are produced via neutrino-nucleus scattering, followed by their decay to the dark gauge boson,  which  in turn gives rise to  
electron-like events. This mechanism provides an excellent fit to
MiniBooNE energy spectra and angular distributions.
\end{abstract}

\maketitle

\emph{Introduction.---}Neutrinos have been connected to anomalies in experimental data since
their commencement in the realm of Physics. From the problems with
beta decays in the dawn of the XX$^\mathrm{th}$ century, that culminated with the
proposal and subsequent discovery of the first of these remarkable
particles, to the solar and atmospheric neutrino puzzles, that
revealed the phenomenon of neutrino oscillations driven by masses and
mixings, the neutrino road has been full of surprises.  Some, however,
like the 17-keV neutrino~\cite{Simpson:1989cu} or the superluminal
neutrinos~\cite{Adam:2011faa} turned out to be mere bumps on the road
as they were resolved by explanations unrelated to new
physics.  As it happens, one never knows which {\em small clouds}
hovering on the horizon of Physics will eventually vanish and which
will instead ignite a revolution.

Even today some peculiar data anomalies remain unsolved. 
On one hand, there is an  apparent
deficit of $\overline \nu_e$ in short-baseline reactor
experiments~\cite{Mention:2011rk} and of $\nu_e$ in radioactive-source
experiments~\cite{Giunti:2010zu}, both amounting  to a 2.5-3$\sigma$
discrepancy that many believe may be connected to unknown nuclear
physics.  On the other hand, the LSND~\cite{Aguilar:2001ty} and MiniBooNE neutrino
experiments~\cite{AguilarArevalo:2007it,AguilarArevalo:2008rc,AguilarArevalo:2010wv,Aguilar-Arevalo:2013pmq}
have reported an excess of $\nu_e$ and $\overline \nu_e$
charge-current quasi-elastic (CCQE) events in their data.  All these
conundrums have been offered a number of exotic interpretations in the
literature~\cite{Gninenko:2009ks,Bai:2015ztj,Liao:2016reh,Carena:2017qhd,Asaadi:2017bhx},
typically invoking eV sterile neutrinos in schemes easily in tension
with other neutrino data~\cite{Collin:2016rao,Gariazzo:2017fdh,Dentler:2018sju}.

Recently, after 15 years of running, MiniBooNE updated their analysis revealing that the 
excess of electron-like events in the  experiment~\cite{Aguilar-Arevalo:2018gpe},
 consistently observed in the neutrino and antineutrino modes,
is now a 4.8$\sigma$ effect. That makes the MiniBooNE result the most statistically relevant anomaly in the neutrino sector.
The origin of such excess is unclear -- it could be the presence of new physics, or a large background mismodeling.
In this Letter  we propose a phenomenological solution to understand the
MiniBooNE data~\footnote{In principle, the mechanism proposed here could provide an explanation of the LSND anomaly. As we will show, the MiniBooNE excess in our framework is induced by a novel neutral current scattering in which neutrinos up-scatter to heavy neutrinos followed by their decays to a collimated $e^+e^-$ pair. Such scattering could kick out a neutron from Carbon in LSND, and thus provide the key signature in inverse beta decay. However, a reliable analysis  of LSND would require detailed experimental information and is beyond the scope of this manuscript.}.

\emph{Framework.---}We introduce a new sector dark~\footnote{To avoid confusion with the vast literature on sterile neutrino models and numerous variants (see e.g. Refs.~\cite{Albright:1971ze, Kolb:1987qy, Vaitaitis:1999wq, Hannestad:2005ex, Chu:2018gxk, Sirunyan:2018mtv, Abada:2018sfh}), we refer to particles in this sector as dark.} composed by a new vector boson, $Z_{\cal D}$, 
coupling directly solely to a dark neutrino, $\nu_{\cal D}$, 
which mixes with the standard ones as
\begin{equation}
\nu_\alpha = \sum_{i=1}^{3} U_{\alpha i} \, \nu_i + U_{\alpha 4} \, N_{\cal D}\, ,\quad \alpha=e,\mu,\tau,{\cal D},
\label{eq:mix}
\end{equation}
where $\nu_i$ and $\nu_{\alpha}$ are the neutrinos mass and flavor eigenstates, respectively.
The new vector boson will, in general, communicate with the Standard Model (SM)  sector via  
either mass mixing or kinetic mixing. The relevant part of the dark Lagrangian is 
\begin{widetext}
\begin{equation}
{\cal L}_{\cal D} \supset \frac{m^2_{Z_{\cal D }}}{2} \, 
Z_{{\cal D}\mu} Z_{\cal D}^{\mu} + g_{\cal D} Z_{\cal D}^\mu \, 
\overline{\nu}_{\cal D} \gamma_\mu \nu_{\cal D} + e \epsilon \, Z_{\cal D}^\mu \, 
J_\mu^{\rm em}
+ \frac{g}{c_W} \epsilon' \, Z_{\cal D}^\mu \, 
J_\mu^{\rm Z} \, ,
\label{eq:kmix}
\end{equation}
\end{widetext}
 where $m_{Z_{\cal D}}$ is the mass of $Z_{\cal D}$ and $g_{\cal D}$ is the coupling 
in the dark sector, $e$ is the electromagnetic coupling, $g/c_W$ is the $Z$ coupling in the SM, while $\epsilon$ and $\epsilon'$ parametrize the kinetic and mass mixings, respectively. The electromagnetic and $Z$ currents are denoted by $J^{\rm em}_\mu$ and $J^{Z}_\mu$.
For simplicity, we assume the mass mixing between the $Z$ and the 
$Z_{\cal D}$ boson to be negligible. We resort to kinetic mixing
between $B_{\mu \nu}$  and $B_{\mu \nu}^\prime$~\cite{Holdom:1985ag}, 
the SM hypercharge and the dark field strengths, 
as a way to achieve  a naturally small coupling between the $Z_{\cal D}$ 
and the electromagnetic current $J_\mu^{\rm em}$. 
 We will take $m_{N_{\cal D}}> m_{Z_{\cal
     D}}$, so the dark neutrino can decay as $N_{\cal D} \to Z_{\cal   D} + \nu_i$, and $m_{Z_{\cal D}}< 2 \, m_\mu$ so the $Z_{\cal D}$
 can only decay to electrons and light neutrinos.
The dark neutrino decay width into $Z_{\cal D}+\nu'{\rm s}$ is simply
\begin{widetext}
\begin{equation}
\Gamma_{N_{\cal D}\to Z_{\cal D}+\nu'{\rm s}} = \frac{\alpha_{\cal D}}{2}\, 
\vert U_{D 4}\vert^2(1-\vert U_{D 4} \vert^2)\, \frac{m^3_{N_{\cal D}}}{m_{Z_{\cal D}}^2}
\left(1-\frac{m^2_{Z_{\cal D}}}{m^2_{N_{\cal D}}}\right)\left(1+
\frac{m^2_{Z_{\cal D}}}{m^2_{N_{\cal D}}}-2 \frac{m^4_{Z_{\cal D}}}{m^4_{N_{\cal D}}}\right)
\, , 
\label{eq:decaywidth-nu}
\end{equation}
\end{widetext}
while the $Z_{\cal D}$ decay width into $e^+e^-$ and light neutrinos are, respectively,
\begin{equation}
\Gamma_{Z_{\cal D}\to e^+e^-} \approx \frac{\alpha \,\epsilon^2}{3}\, m_{Z_{\cal D}}\, ,
\label{eq:zee}
\end{equation}
and
\begin{equation} 
\Gamma_{Z_{\cal D}\to \nu\nu} = \frac{\alpha_{\cal D}}{3} \, \left( 1-\vert U_{D4}\vert^2 \right)^2 
\, m_{Z_{\cal D}} \,. 
\label{eq:znunu}
\end{equation}
We observe that as long as $\alpha\epsilon^2 \gg \alpha_{\cal D}(1-\vert U_{D4}\vert^2)^2$, 
$Z_{\cal D}$ will mainly decay into $e^+e^-$ pairs.

For simplicity, we focus on the case in which both $N_{\cal D}$ and $Z_{\cal  D}$ decay promptly.
Taking the typical energy $E_{N_{\cal D}},E_{Z_{\cal D}} \sim $ 1 GeV, 
and assuming for simplicity 
$\vert U_{e4}\vert^2, \vert U_{\tau 4}\vert^2 \ll \vert U_{\mu 4}\vert^2$,
we can estimate
 $\gamma \, c \, \tau_{N_{\cal D}} \approx 4\times
 10^{-8}m^2_{Z_{\cal D}}[\rm MeV^2]/(m^4_{N_{\cal D}}[\rm MeV^4] \, \alpha_{\cal D} \, \vert U_{\mu
   4}\vert^2)$ cm 
and  $\gamma \, c \, \tau_{Z_{\cal D}} \approx
 6\times 10^{-8}/(m^2_{Z_{\cal D}}[\rm MeV^2] \, \alpha  \epsilon^2)$ cm. So for 
 $\alpha_{\cal D} \sim 0.25$, $\vert U_{\mu 4} \vert^{2}\sim 10^{-8}$ and 
 $\alpha  \epsilon^2 \sim 2 \times 10^{-10}$, 
 $5~{\rm MeV}\lesssim m_{Z_{\cal D}} < m_{N_{\cal D}} $ would guarantee prompt decay for both particles. 
We will see shortly that $m_{N_{\cal D}}$ and $m_{Z_{\cal D}}$ between 
a few tens to a few hundreds of MeV is exactly what is needed to explain the
 experimental data.

\emph{Analysis and results.---}The MiniBooNE experiment is a pure mineral oil (CH$_2$) detector
located at the Booster Neutrino Beam line at Fermilab. The Cherenkov
and scintillation light emitted by charged particles traversing the
detector are used for particle identification and neutrino energy
reconstruction, assuming the kinematics of CCQE scattering.  MiniBooNE
has observed an excess of $381 \pm 85.2$ ($79.3 \pm 28.6$)
electron-like events over the estimated background in neutrino
(antineutrino) beam configuration in the energy range $200 <
E_\nu^{\rm rec}/\rm MeV < 1250$ corresponding to $12.84 \times
10^{20}$ ($11.27 \times 10^{20}$) protons on
target~\cite{Aguilar-Arevalo:2018gpe}.

\begin{figure}[t]
\includegraphics[width=0.3\textwidth]{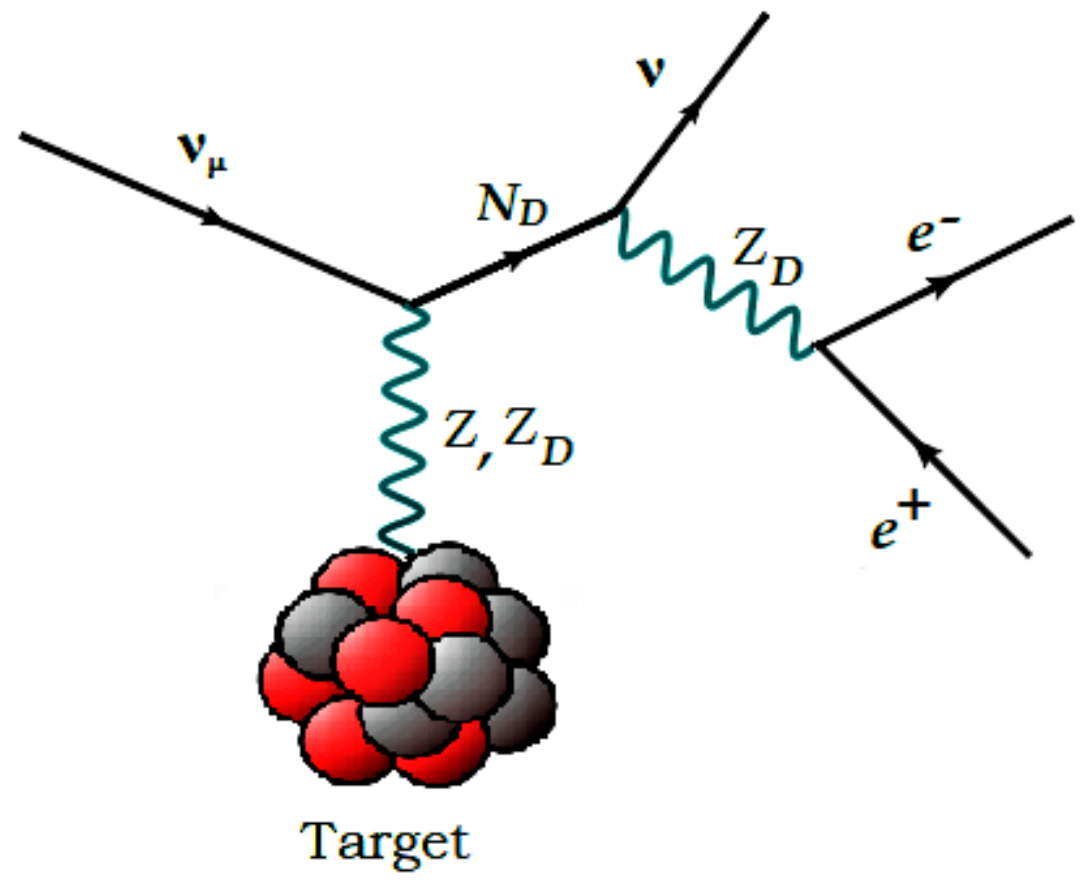}
\caption{\label{fig:feyn} 
Contributions to the cross section that in  our model  
gives rise to MiniBooNE's excess of electron-like events.}
\end{figure}

Our proposal to explain MiniBooNE's low energy excess  from the production and decay of 
a dark neutrino relies on the fact that MiniBooNE cannot distinguish a collimated   $e^+ e^-$
pair from a single electron. Muon neutrinos produced in the beam would up-scatter on the
mineral oil to dark neutrinos, which will subsequently lead to $Z_{\cal D}\to e^+e^-$ as  
shown schematically in Fig.~\ref{fig:feyn}. If $N_{\cal D}$ is light enough, this up-scattering 
in CH$_2$ can be coherent, enhancing the cross section. To take that into account, we 
estimate the up-scattering cross section to be
\begin{equation}
  \frac{d\sigma_{\rm total}/dE_r}{\rm proton} = \frac{1}{8}F^2(E_r)\frac{d\sigma_{\rm C}^{\rm coh}}{dE_r}
     +\left(1-\frac{6}{8}F^2(E_r)\right)\frac{d\sigma_p}{dE_r},
\end{equation}
where $F(E_r)$ is the nuclear form factor~\cite{Engel:1991wq} for
Carbon, while $\sigma_{\rm C}^{\rm coh}$ and $\sigma_p$ are the elastic
scattering cross sections on Carbon and protons, which can be easily
calculated. For Carbon, $F(E_r)$ is sizable up to proton recoil
energies of few MeV.

To obtain the spectrum of events, a simplified model was implemented
in FeynRules~\cite{Alloul:2013bka} in which Carbon and protons were
taken to be an elementary fermion and events were generated in
MadGraph5~\cite{Alwall:2014hca}.  Since MiniBooNE would interpret
$Z_{\cal D}\to e^+e^-$ decays as electron-like events, the
reconstructed neutrino energy would be incorrectly inferred by the
approximate CCQE formula (see e.g. Ref.~\cite{Martini:2012fa})
 \begin{equation}
  E_\nu^{\rm rec} \simeq \frac{m_p \, E_{Z_{\cal D}}}{m_p-E_{Z_{\cal D}}(1-\cos\theta_{Z_{\cal D}})},
\end{equation}
where $m_p$ is the proton mass, and $E_{Z_{\cal D}}$ and $\theta_{Z_{\cal D}}$ are
the dark $Z_{\cal D}$ boson energy and its direction relative to the
beam line.  The fit to MiniBooNE data was then performed using the $\chi^2$ function from the
collaboration official data release~\cite{Aguilar-Arevalo:2018gpe}, which includes the $\nu_\mu$ and $\bar\nu_\mu$ disappearance data,
re-weighting the Montecarlo events by the ratio of our cross section
to the standard CCQE one, and taking into account the wrong sign contamination 
from Ref.~\cite{AguilarArevalo:2008yp}. Note that the official covariance matrix includes spectral data in electron-like and muon-like events for both neutrino and antineutrino modes.

\begin{figure}[t]
\includegraphics[width=0.51\textwidth]{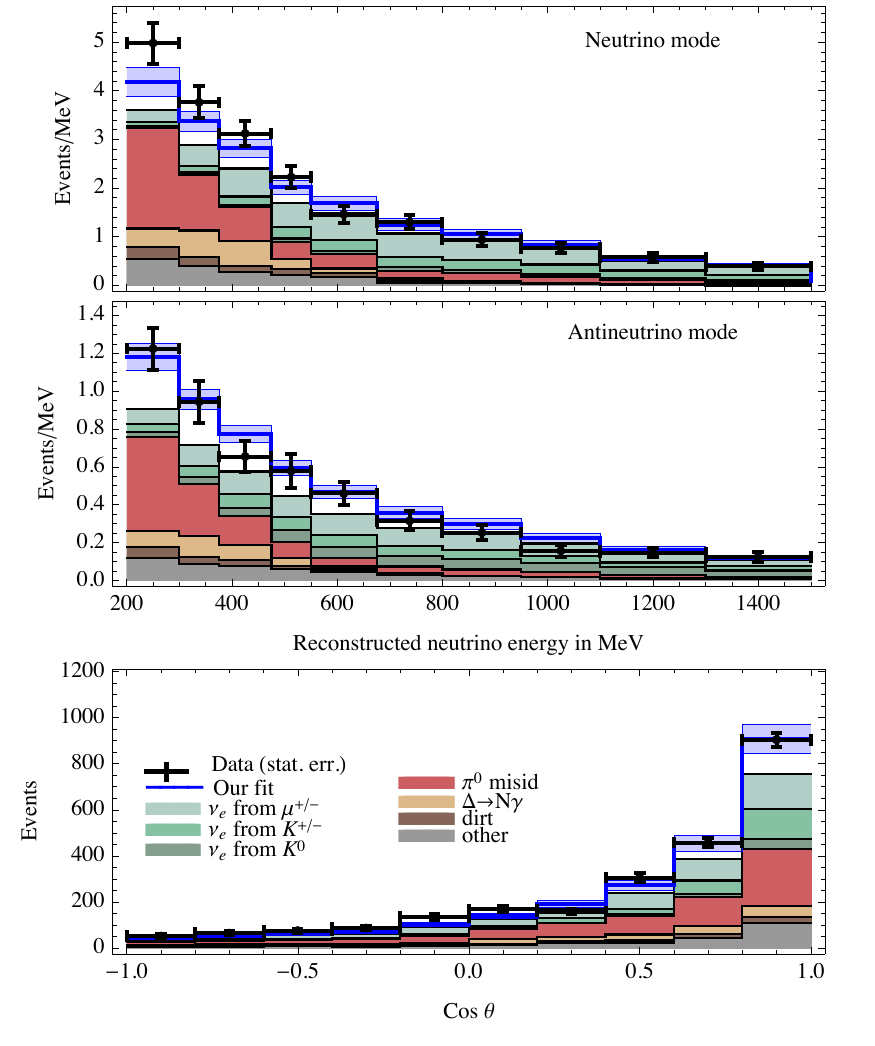}
\caption{\label{fig:spectra} 
The MiniBooNE electron-like event data~\cite{Aguilar-Arevalo:2018gpe} in 
the neutrino (top panel) and antineutrino (middle panel) modes 
 as a function of $E_\nu^{\rm rec}$, as well as the $\cos \theta$ 
distribution (bottom panel) for the neutrino data. Note that the data points have only statistical uncertainties, while the systematic uncertainties from the background are encoded in the light blue band..
The predictions of our benchmark point  $m_{N_{\cal D}}= 420$ MeV, $m_{Z_{\cal D}}=30$ MeV, 
$\vert U_{\mu 4}\vert^2=9\times 10^{-7}$, $\alpha_{\cal D}=0.25$ and 
$\alpha\, \epsilon^2= 2 \times 10^{-10}$ are also shown as the blue lines.}
\end{figure}

In Fig.~\ref{fig:spectra} we can see the  electron-like event distributions, 
including all of the backgrounds, as reported by MiniBooNE. 
We clearly see the event excess reflected in all of them. 
The neutrino (antineutrino) mode data as a function of $E_\nu^{\rm rec}$ is displayed
on the top (middle) panel. 
The corresponding  predictions of our model, for the benchmark point 
$m_{N_{\cal D}}= 420$ MeV, $m_{Z_{\cal D}}= 30$ MeV, $\vert U_{\mu
  4}\vert^2= 9\times10^{-7}$, $\alpha_{\cal D}=0.25$ and 
$\alpha \epsilon^2= 2\times 10^{-10}$, are depicted as the blue lines. The light blue band reflects an approximated systematic uncertainty from the background estimated from Table I of Ref.~\cite{Aguilar-Arevalo:2018gpe}.
On the bottom panel we show the $\cos \theta$ distribution 
of the electron-like  candidates for the neutrino data, 
 as well as the distribution for $\cos \theta_{Z_{\cal D}}$ 
for the benchmark point (blue line).
The $\cos \theta$ distribution of the electron-like candidates in the 
antineutrino data is similar and not shown here and our model 
is able to describe it comparably well. 
We remark that our model prediction is in extremely good agreement 
with  the experimental data. In particular, our fit to the data
is better than the fit under the electronVolt sterile neutrino oscillation 
hypothesis~\cite{Aguilar-Arevalo:2018gpe} if one considers the constraints from other oscillation experiments. We find a best fit with $\chi^2_{bf}/{\rm dof}=33.2/36$, while the background only hypothesis yields $\chi^2_{bg}/{\rm dof}=63.8/38$, corresponding to a $5.2\sigma$ preference for our model.

In our framework, as the dark boson decays dominantly to charged
fermions, the constraints on its mass and kinetic mixing are
essentially those from a dark photon~\cite{Ilten:2018crw}. In the mass
range $20\sim 60$~MeV, the experiments that dominate the phenomenology
are beam dump experiments and  NA48/2. 
Regarding the dark neutrino, the constraints are similar but weaker
than in the heavy sterile neutrino scenario with non-zero
$|U_{\mu4}|^2$~\cite{Atre:2009rg, deGouvea:2015euy}. Since $N_{\cal
D}\to\nu e^+ e^-$ is prompt, limits from fixed target experiments like
PS191~\cite{Bernardi:1987ek}, NuTeV~\cite{Vaitaitis:1999wq}, BEBC~\cite{CooperSarkar:1985nh}, FMMF~\cite{Gallas:1994xp} and CHARM II~\cite{Vilain:1994vg} do not apply. Besides, $W\to\ell
N\to \ell\nu e^+e^-$ in high energy colliders can constrain
$|U_{\mu4}|^2>{\rm few}\times 10^{-5}$ for $m_{N_{\cal D}}>\mathcal{O}({\rm
GeV})$~\cite{Sirunyan:2018mtv}. Finally, we do not expect any significant constraints from the MiniBooNE beam dump run~\cite{Aguilar-Arevalo:2018wea} due to low statistics.

\begin{figure}[t]
\includegraphics[width=0.5\textwidth]{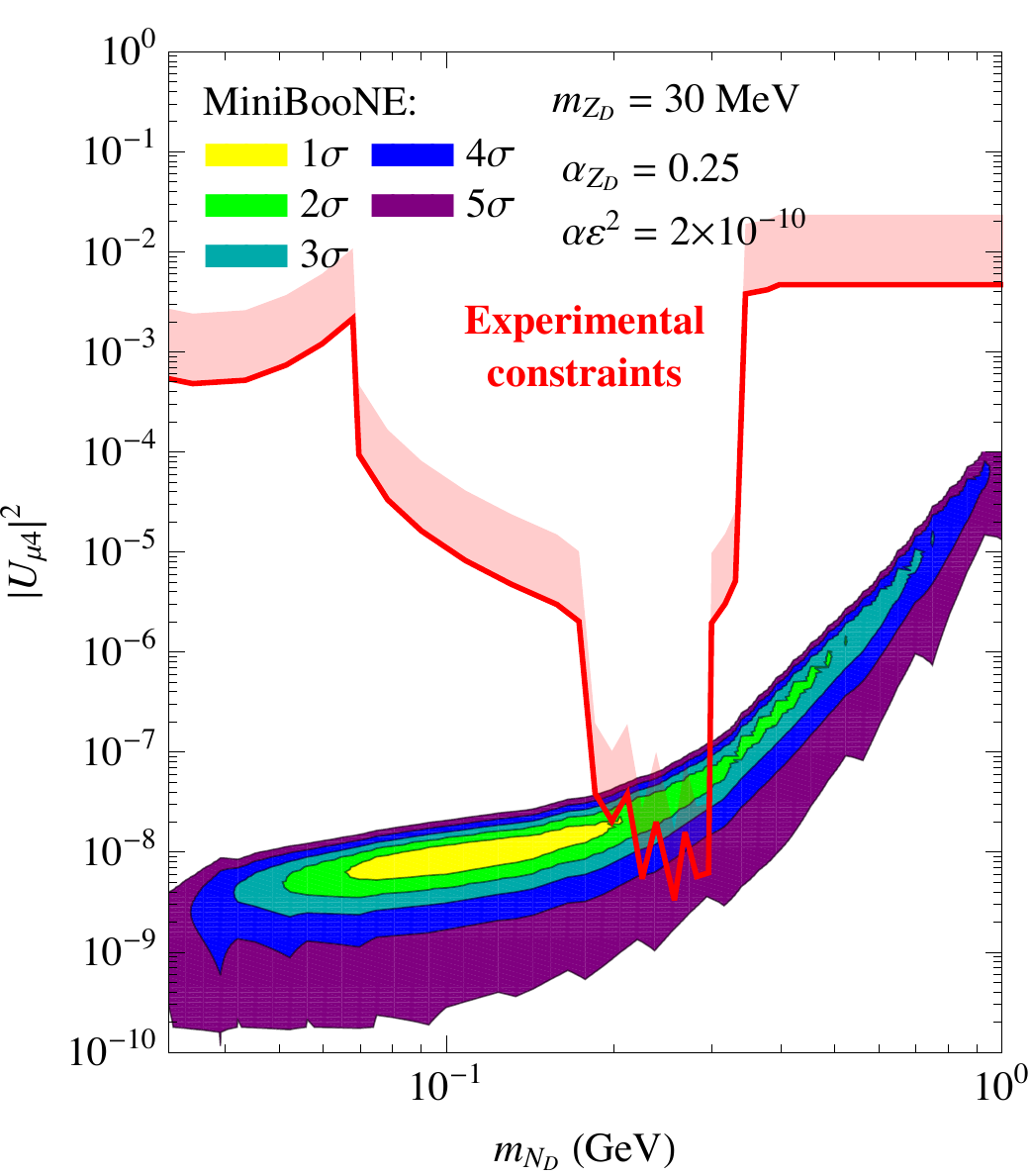}
\caption{\label{fig:mp} 
Region of our model in the $\vert U_{\mu 4}\vert^2$ versus $m_{N_{\cal D}}$ plane 
satisfying MiniBooNE data at 1$\sigma$ to 5$\sigma$ CL, for 
the hypothesis $m_{Z_{\cal D}} =  30$~MeV, $\alpha_{Z_{\cal D}}=0.25$ and
$\alpha \epsilon^2 = 2 \times 10^{-10}$. The region above the red curve is excluded at 
99\% CL by meson decays, the muon decay Michel spectrum and lepton 
universality~\cite{Atre:2009rg,deGouvea:2015euy}.}
\end{figure}

In Fig.~\ref{fig:mp} we see the region in the plane $\vert U_{\mu 4}\vert^2$ versus $m_{N_{\cal D}}$
consistent with MiniBooNE data at 1$\sigma$ to 5$\sigma$ CL,
for the exemplifying hypothesis $m_{Z_{\cal D}} = 30$~MeV, $\alpha_{Z_{\cal D}}=0.25$ and
$\alpha \epsilon^2 = 2 \times 10^{-10}$. 
Other values of these parameters can also provide good agreement with the data.
We also show the combined non-oscillation bounds from  meson decays, muon decay Michel 
spectrum and lepton universality compiled in Refs.~\cite{Atre:2009rg,deGouvea:2015euy}, 
which exclude the region above the red line. 
The ship hull shape region can be divided in two parts:
a high mixing region at $\vert U_{\mu 4}\vert^2 \sim 10^{-4}-10^{-8}$, corresponding 
to $m_{N_{\cal D}} \gtrsim 300$ MeV, and a low mixing region for $\vert U_{\mu 4} \vert^2 
\lesssim 10^{-8}$ and $m_{N_{\cal D}} \lesssim 200$ MeV. The latter seems to be 
favored by spectral data.
As a side remark, we have checked that the typical opening angle $\theta_{e^+e^-}$ of the $e^+e^-$ pair satisfy $\cos\theta_{e^+e^-}>0.99$, ensuring that MiniBooNE will identify these events as electron-like.

The MicroBooNE experiment at Fermilab~\cite{Antonello:2015lea} is currently investigating 
the low energy excess of electron-like events observed by MiniBooNE. 
They can distinguish electrons from photon conversions into a $e^+e^-$ pair  by 
their different ionization rate at the beginning of their trajectory in the 
liquid argon detector. 
In addition our framework allows for the possibility of the experimental observation of  the 
 $K_L \to \nu_{\cal D} \nu_{\cal D}$, via off-shell $Z_{\cal D}$ exchange, by the KOTO or NA62 experiments 
as ${\cal B}(K_L \to \nu_{\cal D} \nu_{\cal D})$ can go up to ${\cal O}(10^{-10})$ for $m_{N_{\cal D}}< m_K$~\cite{Abada:2016plb}.

We also have inquired into the possible effects of $N_{\cal D}$ and $Z_{\cal D}$
on  oscillation experiments. 
While low energy sources, such as the sun or 
nuclear reactors, do not have  enough energy to produce  these particles, 
they could be, in principle, produced in higher energy oscillation experiments.
Typically $\nu_\mu$ and $\overline \nu_\mu$ beams in accelerator 
neutrino experiments have an insurmountable $\mathcal{O}(1\%)$ contamination 
of $\nu_e+\overline \nu_e$, and atmospheric neutrinos have a 
large $\nu_e$ and $\overline \nu_e$ component. 
While Cherenkov detectors, like Super-Kamiokande, cannot distinguish 
between electrons and photons, detectors like MINOS, NO$\nu$A or T2K
would have a hard time to see any signal over their neutral current
contamination. That is particularly relevant  at lower energies where one would expect 
the signal of new physics to lay.

In a different note, we do not foresee any issues with cosmological data, as the particles in the dark sector decay too fast to affect Big Bang Nucleosynthesis, and the $\nu-\nu$ self-interactions are too small to change neutrino free streaming. Supernova cooling would not constrain the model, as the $Z_{\cal D}$ is trapped due to the large kinetic mixing.

Finally, one may wonder if the phenomenological approach we propose here can 
arise in a UV-complete anomaly free model. We have checked that such realization is possible as follows. A gauge $U(1)_{\cal D}$ symmetry, under which the only charged fermions are the dark neutrinos, protects neutrino masses from the standard Higgs mechanism. An enlarged scalar sector is called upon to ensure non-zero neutrino masses, naturally leading to $\nu-N_{\cal D}$ mixing, as well as the mass of the dark gauge boson. In this realization, both kinetic and mass mixing are unavoidable, but typically small.
The model naturally connects neutrino masses with the new interaction~\cite{Bertuzzo:2018ftf}. 
We will explore the rich phenomenology of this model in detail elsewhere.

\emph{Conclusion.---}We have  shown that the low energy excess observed by MiniBooNE can by explained by a light dark sector to which neutrinos are a portal. The framework is elegant and no tuning is needed to fit the excess.
We find an excellent agreement with spectral and angular data distributions, in both neutrino and antineutrino modes. This solution is consistent with all current experimental data and can be probed by Liquid Argon detectors in the near future.

\begin{acknowledgments}
We are grateful to Roni Harnik, William Louis, Xiao Luo, Ornella Palamara, Stefan Prestel for useful discussions.
This work was partially supported by Funda\c{c}\~ao de Amparo \`a
Pesquisa do Estado de S\~ao Paulo (FAPESP) and Conselho Nacional de
Ci\^encia e Tecnologia (CNPq).  
R.Z.F. is greatful for the hospitality of the Fermilab Theory Group 
during the completion of this work.
The work of S.J. is supported in part by the US Department of Energy Grant (DE-SC0016013) and the Fermilab 
Distinguished Scholars Program. Fermilab is operated by the Fermi Research Alliance, LLC under contract No. DE-AC02-07CH11359 with the United States Department of Energy. 
This project has received support from the
European Union's Horizon 2020 research and innovation programme under
the Marie Sk\l{}odowska-Curie grant agreements No 690575
(InvisiblesPlus) and No 674896 (Elusives).
\end{acknowledgments}

\bibliography{bjmz}

\end{document}